\documentstyle[11pt,newpasp,twoside,epsf]{article}
\def\edcomment#1{\iffalse\marginpar{\raggedright\sl#1\/}\else\relax\fi}
\reversemarginpar

\begin{document}
\title{Galactic Winds in Starburst Irregular Galaxies}
 \author{Simone Recchi, Francesca Matteucci}
\affil{Dipartimento di Astronomia, Universit\`a di Trieste\\
Via G.B. Tiepolo 11, 34131 Trieste, Italy}
\author{Annibale D'Ercole}
\affil{Osservatorio Astronomico di Bologna\\
Via Ranzani 1, 44127 Bologna, Italy}

\begin{abstract}

In this paper we present some results of numerical simulations
concerning the development of galactic winds in starburst galaxies. In
particular, we focus on a galaxy similar to IZw18, the most metal-poor
galaxy locally known. We compute the chemo-dynamical evolution of this
galaxy, considering the energetic input and the chemical yields
originating from Supernovae (SNe) of Type II and Ia and from
intermediate-mass stars. We consider both single, instantaneous
starburst and two starburst separated by a quiescent period. In all
considered cases a metal enriched wind develops and in particular the
metals produced by Type Ia SNe are ejected more efficiently than the
other metals. We suggest that two bursts of star formation, the first
being weaker and the last having an age of some tenth of Myr, can
satisfactorily reproduce the abundances and abundance ratios found in
literature for IZw18.

\end{abstract}

\section{Introduction}

Many Dwarf Irregular Galaxies (DIG) are known to be in a starburst
phase, or are believed to have experienced periods of intense star
formation in the recent past. These galaxies are generally called Blue
Compact Dwarfs (BCD). Energetic events associated with star formation
(stellar winds and Supernova explosions) may sweep the interstellar
gas out of the region actively forming stars, thus creating galactic
winds. Observational evidences of outflows have been found recently in
many edge-on starburst galaxies, like NGC1705, NGC1569 and NGC3628.
Martin (1998) found that large expanding supershells are common
byproducts of massive star formation in dwarf galaxies.

Both dynamical and chemical simulations of these galaxies have
suggested the existence of a `differential galactic wind', in the
sense that, after a starburst event, these objects would loose mostly
metals (Pilyugin 1992, 1993; Marconi, Matteucci \& Tosi 1994; MacLow
\& Ferrara 1999; D'Ercole \& Brighenti 1999). However in none of these
studies, detailed chemical and dynamical evolution was taken into
account at the same time.

Among BCD, the galaxy IZw18 (the most metal-poor galaxy locally known)
constitutes the best candidate for a truly ``young'' galaxy. However,
there is still a debate in literature on whether IZw18 is experiencing
the first burst of star formation or not. Evolutionary population
synthesis models by Mas-Hesse \& Kunth (1999) show that the present
burst is very young (between 3 and 13 Myr) and the contribution of
older stars, if any, is negligible. Dynamical arguments (Martin 1996)
suggest a single burst with an age between 15 and 27 Myr. Recent Color
Magnitude Diagram (CMD) studies of IZw18, both in the optical (Aloisi,
Tosi \& Greggio 1999; hereafter ATG) and in the infrared (\"Ostlin
2000), revealed the presence of an underlying older population, with
an age of some 10$^8$ yr. Legrand (2000) and Legrand et al. (2000)
proposed instead a low and continuous star formation regime for IZw18.

We study, through numerical simulations, the dynamical and chemical
evolution of a gas-rich dwarf galaxy whose structural parameters
resembles IZw18. We consider single instantaneous starburst or a
couple of starburst separated by a quiescent period.  We include
effects (both energetical and chemical) of Type II and Type Ia SNe in
a detailed way. The aim of this work is to test the `differential
wind' hypothesis with an hydrodynamical approach and to find
constraints for the number and for the age of the starburts in IZw18.

\section{The model}

We consider a gaseous component in hydrostatic isothermal equilibrium
with the centrifugal force and a potential well. The potential well is
the sum of a quasi-isothermal dark halo ($M_{dark} \simeq 6.5 \cdot
10^{8}$ M$_\odot$) and an oblate King profile. The mass of gas inside
the galactic region (an ellipsoid with dimensions 1 Kpc $\times$ 
730 pc) is $\sim 1.7 \cdot 10^7$ M$_\odot$. 

To describe the evolution of the ISM, we adopt a 2-D hydrocode, with
source terms describing the rate of mass and energy return from the
starbursts, taking into account SNe of Type II and Ia and low and
intermediate-mass stars (IMS). By using passively evolving tracers, we
are able to follow the evolution, in space and time, of some chemical
elements of particular astrophysical interest. The production of these
elements are obtained following the nucleosynthesis prescriptions from
various authors: Woosley \& Weaver (1995) for SNeII, Nomoto,
Thielemann \& Yokoi for SNeIa and Renzini \& Voli (1981), case R, or
van den Hoek \& Groenewegen (1997), case V, for IMS. See Recchi,
Matteucci \& D'Ercole (2001) for more details about model
prescriptions.

In the single-burst model, the mass of stars formed is $M_\star = 6
\cdot 10^5$ M$_\odot$, whereas in the two-burst model we consider a
weaker first burst (the mass of stars is $10^5$ M$_\odot$). After 300
Myr (model M300) or 500 Myr (model M500) we consider that 10\% of cold
gas inside a central region (a sphere of 200 pc of radius) is turned
into stars. The initial abundances of this second stellar generation
are simply the metallicity of the cold gas in the central region at
the onset of the burst. According to the results of ATG, we consider
also a flatter IMF (model M300F), with a slope $x=0.5$. More details
about these two-burst models are in Recchi et al. (2001b).

\vspace{6cm}
The efficiency of SN heating is assumed to be very low for Type II
SNe. In particular, following results of Bradamante, Matteucci \&
D'Ercole (1998), we consider a thermalization efficiency of
$\eta_{II}=0.03$ for the single-burst model and $\eta_{II}=0.05$ for
the two-burst model, namely only 3\% (or 5\%) of the explosion energy
is able to thermalize the ISM, while the rest is radiated away. Type
Ia SNe instead explode in an already heated and diluted medium, thus
their thermalization efficiency is assumed to be $\eta_{Ia}=1$. There
is debate in literature about the correct value of thermalization
efficiency. In particular, Strickland \& Stevens (1999) assume a value
$\eta=1$ (see also the contribution of Strickland in this volume). The
reason of our choice is that the number of SNeII exploding in our
model is rather low, thus the time interval between single explosions
($\sim 6 \cdot 10^4$ yr) is larger than the typical cooling
time-scale, thus remnant of Type II SNe evolve as single SNR. In
addition, we tried to run simulations with an efficiency $\eta=1$ also
for Type II SNe and the galaxy is quickly devoided of gas, thus the
actual gas content in IZw18 rules out the possibility of an high
$\eta_{II}$.

\section{Results}
\subsection{Dynamical results}
\subsubsection{Single-burst model}

\begin{figure}
\vspace{1cm}
\plotfiddle{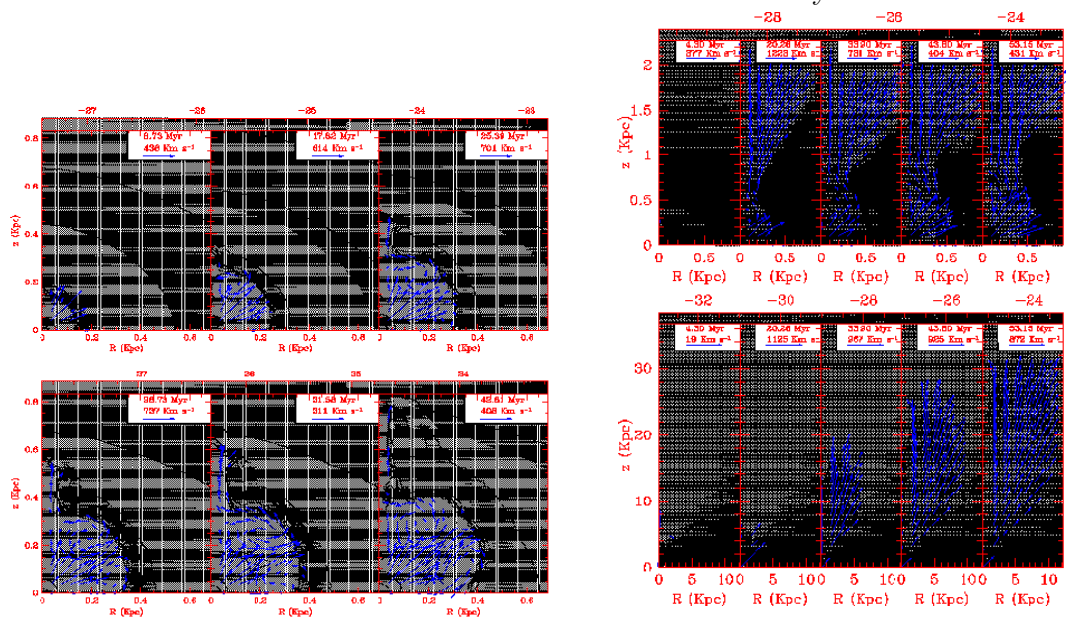}{200pt}{0}{39}{38}{-220pt}{-40pt}
\caption{Density contours and velocity field for the single-burst
model (left panels) and two-burst model M300 (right panels) at
different ephocs (age of the burst is labelled inside each panel). The
density scale (logarithmic) is labelled in the strip on top of the
figures. For what conserns model M300 (right panels), upper panels are
a zoom in the central regions of what shown in the lower panels.}
\end{figure}

\vspace{-4cm}
\begin{figure}

\plotfiddle{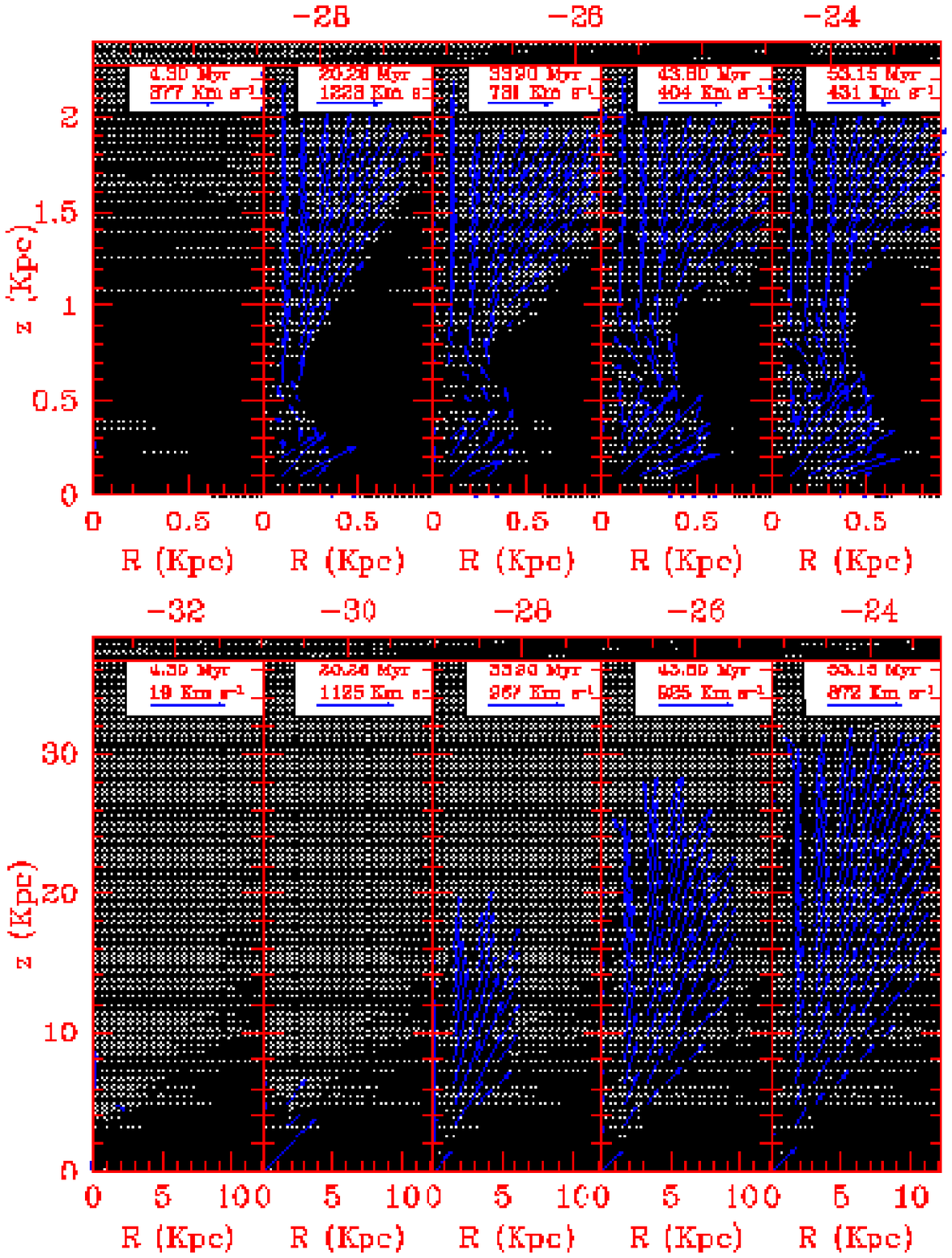}{-270pt}{0}{42}{42}{-20pt}{85pt}

\end{figure}
\vspace{4cm}

Owing to the energy released by Type II SNe, a galactic wind
develops. It expands faster along the z direction, where the ISM
density gradient is steeper. SNeII activity lasts for only 29 Myr (the
lifetime of a 8 M$_\odot$ star), then is replaced by a weaker SNIa
wind, not strong enough to sustain the galactic outflow (Fig. 1).
After $\sim$ 300 Myr, the expanding ISM is diluted enough and the hot
bubble finally breaks out through a funnel. Most of the SNeII ejecta
reamins locked into the cold and dense shell, whereas metals ejected
by SNeIa are easily channelled along the funnel. Iron, mostly produced
by Type Ia SNe, is thus easily lost by the galaxy and the
[$\alpha$/Fe] ratios ouside the galaxy are lower than inside (see
section 4).  Owing to the low evolution of the superbubble, the
internal cavity becomes soon radiative (i.e. radiates an energy
comparable to the thermal energy content of the shocked wind) thus
after $\sim$ 10 Myr most of metals are in a cold phase.

\begin{figure}
\plotfiddle{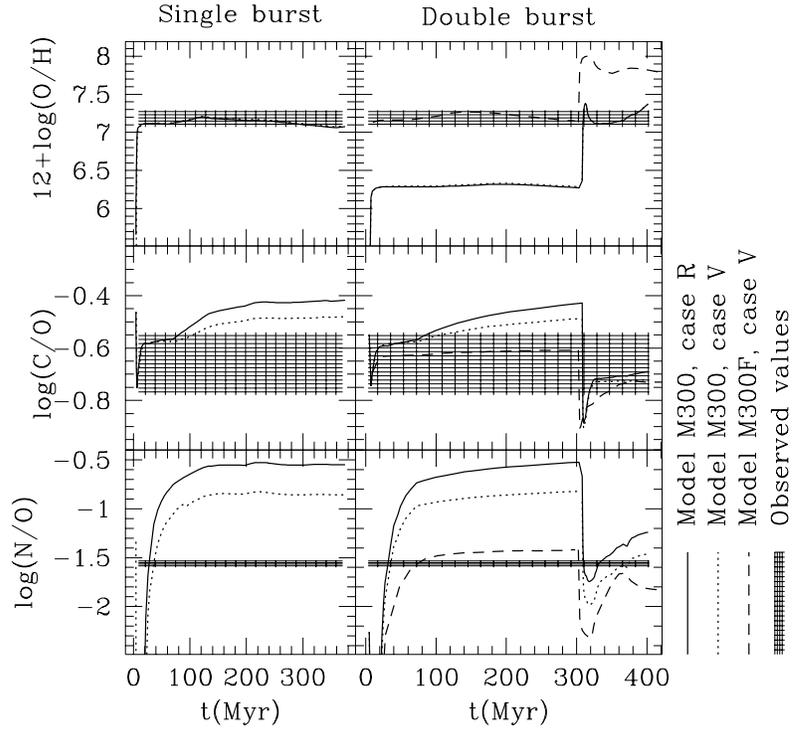}{300pt}{0}{55}{55}{-140pt}{-70pt}
\vspace{-1cm}
\caption{Evolution of O, C/O and N/O for the single-burst model and
the models M300 and M300F. Left panels represent the evolution of 
the single-burst model with RV81 yields (solid line) and VG97 
(dotted line). In the right panels the evolution of model M300 with 
RV81 yields (solid line) and VG yields (dotted lines), together with 
model M300F, VG97 yields (dashed lines) are shown}
\end{figure}

\subsubsection{Two-burst model}
Owing to the low luminosity of the first burst, after $\sim$ 300 Myr a
galactic wind still does not develop. For the model M300 a cold, dense
shell of dimensions 200 $\times$ 100 pc forms and outside this region
the ISM is almost unperturbed. The impact of the second generation of
stars on the ISM dynamics is rather vigorous. Already after $\sim$ 30
Myr after the onset of the second burst a breakout occurs (see Fig. 1)
and the gas produced during this second burst is easily lost along the
galactic chimney.

The hypothesis of a differential wind is substantially confirmed in
these simulations: metals are ejected more easily than pristine
ISM. Also for the two-burst model the ejecta of Type Ia SNe are lost
more easily, but this effect is less evident compared to single-burst
model. The consequence of this selective losses of metals is that
[$\alpha$/Fe] ratios outside the galaxy are lower than inside.

\subsection{Chemical results}

The evolution of Oxygen abundance and C/O and N/O ratios for the
single-burst model, cases R and V (left panels) and for the two-burst
models M300, cases R and V and M300F, case V only (right panels) are
shown in Fig. 2. Shaded areas represents the observed values found in
literature for IZw18

Single-burst model reproduces the observed abundances of IZw18 only
for a very short time, at an age of $\sim$ 31 Myr. After this time,
N/O ratio begins to increase over the permitted range, owing to the N
produced by intermediate-mass stars. Model M300 is able to reproduce
observed abundances of IZw18 for a wider range of times: between 25
and 40 Myr after the onset of the second burst (case R) and between 50
and 70 Myr (case V). The M300F model produces too much oxygen during
the first burst of star formation and does not fit the observed
abundance ratios, unless the age of the second burst is extremely
short (around 4 Myr). Model M500 (not shown here) is able to fit 
abundances found in literature for an evolutionary time between 
40 and 80 Myr.

\section{Conclusions}

Our main conclusions can be summarized as follows:

\begin{itemize}

\item
energetic events associated with the starbursts, are able to trigger a
galactic wind and the metals produced in the burst leave the galaxy
more easily thaan the unprocessed gas.

\item
In particular, the ejecta of Type Ia SNe are lost more efficiently
than Type II SNe, because this kind of explosions occur in a hot and
rarefied medium. the consequence is that [$\alpha$/Fe] ratios outside
the galaxy are lower than inside. This effect is more evident in the
single-burst model.

\item 
Single-burst model reproduces the observed abundances in IZw18 after
$\sim$ 31 Myr, while for the two-burst model we obtain agreement with
the data found in literature for wider ranges of time, when the second
starburst has an age of some tenth of Myr, depending on the adopted
model and nucleosynthesis prescriptions.

\item
The classical Salpeter IMF should be preferred over a flatter one
which would predict a too high oxygen abundance.

\item
Finally, we can suggest that a first, weak burst of star formation,
occurred more than 300 Myr ago, followed by a stronger one, having an
age of some tenth of Myr, with a Salpeter IMF, best reproduces the
properties of IZw18.
\end{itemize}

\end{document}